\documentclass[]{RAA}
\usepackage{graphicx,times}
\usepackage{natbib}

\begin{document}

\title{Pulsating magneto-dipole radiation of a quaking neutron star
powered by energy of Alfv\'en seismic vibrations}

\volnopage{{\bf 2011} Vol.{\bf 11} No.{\bf 9},  preprint}
\setcounter{page}{1}

\author{S.I.Bastrukov
 \inst{1,2}
 \and J.W.Yu
   \inst{1}
\and R.X.Xu
  \inst{1}
   \and I.V.Molodtsova
  \inst{2}
  }

\institute{State Key Laboratory of Nuclear Physics and Technology,
School of Physics, Peking University, Beijing 100871, China\\
\and
Joint Institute for Nuclear Research, 141980 Dubna, Russia.\\
\vs \no
{\small
Received[year][month][day];accepted[year][month][day] } }

\abstract{
We compute the characteristic parameters of magnetic dipole radiation of a neutron star undergoing torsional seismic vibrations under the action of Lorentz restoring force about axis of a dipolar magnetic field  experiencing decay. 
After brief outline of general theoretical background of the model of vibration powered neutron star, we present 
numerical estimates of basic vibration and radiation characteristics, such as 
the oscillation frequency, lifetime, luminosity of radiation, and investigate 
their time dependence upon magnetic field decay. 
The presented analysis suggests that gradual decrease in  
frequencies of pulsating high-energy emission detected from a handful 
of currently monitored  AXP/SGR-like X-ray sources can be explained as being produced by vibration powered magneto-dipole radiation of quaking magnetars.
 }

\authorrunning{S.I.Bastrukov, J.W.Yu, R.X.Xu \& I.V.Molodtsova}
\titlerunning{Vibration powered neutron star}
\maketitle

\section{Introduction}
 The last two decades have seen an increasing interest in the 
 Soft Gamma Repeaters (SGRs) and Anomalous X-ray Pulsars -- commonly 
 referred to as magnetars (Duncan \& Thompson 1992), the seismic and radiative 
 activity of which is fairly different from that of rotation powered radio pulsars (e.g.  
 Kouveliotou 1999, Harding 1999,  Woods \& Thompson 2006, 
 Mereghetti 2008, Qiao, Xu \& Du 2010). The most popular idea is that 
 energy supply of long-periodic pulsating radiation of these fairly young neutron stars 
 comes from process involving decay of ultra strong magnetic field. 
 One of such processes could be magneto-mechanical vibrations 
 driven by forces of magnetic-field-dependent stresses (Bastrukov et al. 2002).
 In the development of this line of argument, particular attention has been given to 
 torsion vibrations of perfectly conducting stellar matter about magnetic axis of the star under the action of magnetic Lorentz force with focus on discrete frequency spectra of toroidal Alfv\'en mode ($a$-mode).  In the past, the standing-wave regime 
 of such vibrations has been subject of several investigations 
 (Ledoux \& Walraven 1958). In works (Bastrukov et al. 2009a, 2009b, 2010) focus was made on the non-investigated regime of node-free vibrations in 
 in the static (time-independent) field. The prime purpose of these latter works
  was to get some insight into difference between spectra of discrete frequencies
  of toroidal $a$-modes in neutron star models having one and the same
  mass  $M$ and radius $R$, but different shapes of constant-in-time poloidal
  magnetic fields. By use of the Rayleigh energy method, it was found that each 
  specific
  form of spatial configuration
  of static magnetic field  about axis of which the neutron star matter undergoes
  node-free differentially rotational oscillations is uniquely reflected in the discrete frequency spectra
  by form of dependence of frequency upon overtone $\ell$ of vibrations.
  The subject of present our study is radiative activity of quaking neutron star 
  powered by energy of Alfv\'en vibrations in its own magnetic field experiencing 
  decay. Part of this project 
  has been reported in recent workshops and conferences and in
  short paper (Bastrukov et al. 2011). Taking this into account, in this article 
  only a brief overview of theoretical background of the model is given. 
  The focus is placed on numerical computation  
  of basic characteristics of vibration and radiation which are of interest in 
  observational search for such objects.

  \section{General background of the vibration powered neutron star model}  
 
 Only a brief outline of this model is given here and more details can be found 
 elsewhere (Bastrukov et al. 2011). The Lorentz-force-driven 
 differentially rotational vibrations of perfectly conducting matter of neutron star about 
 axis of poloidal internal and dipolar external magnetic field evolving in time are  
 properly described in terms of material displacements ${\bf u}$ 
  obeying equation of magneto-solid-mechanics
 \begin{eqnarray}
  \label{e2.1}
 && \rho(r)\, {\ddot {\bf u}}({\bf r},t)=\frac{1}{4\pi}
 [\nabla\times[\nabla\times [{\bf u}({\bf r},t)\times {\bf B}({\bf r},t)]]]\times {\bf B}
 ({\bf r},t)\\
 \label{e2.2}
 && {\dot {\bf u}}({\bf 
   r},t)=[\mbox{\boldmath $\omega$}({\bf r},t)\times {\bf r}],\quad 
   \mbox{\boldmath $\omega$}
  ({\bf r},t)=A_t[\nabla\chi(r)]\,{\dot\alpha}(t).
  \end{eqnarray}
  The field ${\dot {\bf u}}({\bf r},t)$ is identical to that for torsion node-free 
  vibrations restored by  Hooke's force of elastic stresses  (Bastrukov et al.
  2007, 2010)
  with  $\chi({\bf r})=A_\ell\, f_\ell(r)\,P_\ell(\cos\theta)$ where 
  $f_\ell(r)$ is the nodeless function of distance from the star center and 
   $P_\ell(\cos\theta)$ is Legendre polynomial of degree $\ell$ specifying the overtone 
   of toroidal mode. In (\ref{e2.2}), the amplitude ${\alpha}(t)$ is the basic 
   dynamical variable describing time evolution of vibrations which is different for 
   each specific overtone.  The bulk density can be represented in the form   
  $\rho(r)=\rho\phi(r)$, where $\rho$ is the density at the star center and  $\phi(r)$ 
  describes the radial profile of density which can be taken from computations of 
  neutron star structure relying on realistic equations of state accounting for non-
  uniform mass distribution in the star interior (e.g., Weber 1999).
  The central to the subject of our study is the following representation 
   of the time-evolving internal magnetic field ${\bf B}({\bf r},t)=B(t)\,{\bf b}({\bf 
   r})$, where $B(t)$ is the time-dependent intensity 
  and ${\bf b}({\bf r})$ is dimensionless vector-function of the field distribution 
  over the star volume. The gist of the energy variational method of computing 
  frequency of nod-free Alfv\'en vibrations consists in the following 
  separable representation of material displacements ${\bf u}({\bf r},t)={\bf a}({\bf 
  r})\,\alpha(t)$. Scalar product of (\ref{e1.1}) with this form of ${\bf u}$     
  and integration over the star volume leads to equation for amplitude $\alpha(t)$ 
  having the form of equation of oscillator with time-depended spring constant  
\begin{eqnarray}
  \label{e2.3}
  &&{\cal M}{\ddot \alpha}(t)+{\cal K}(t)\alpha(t)=0,\\
  \nonumber
  &&{\cal M}=\rho\,m,\quad  m=\int \phi(r)\,{\bf a}({\bf r})\cdot {\bf a}({\bf r})\,d{\cal V},\quad 
  {\bf a}=A_t\nabla\times [{\bf r}\,f_\ell(r)\,P_\ell(\cos\theta)]\\
   \nonumber
  && {\cal K}=\frac{B^2(t)}{4\pi}\, k,\quad k=\int
 {\bf a}({\bf r})\cdot [{\bf b}({\bf r})\times [\nabla\times[\nabla\times [{\bf a}({\bf
 r})\times {\bf b}({\bf r})]]]]\,d{\cal V}
 \end{eqnarray}
  As for the general asteroseismology of compact objects is concerned, the above
    equations  seems to be appropriate not only for neutron stars but also white
    dwarfs (Molodtsova et al 2010, Bastrukov et all 2010a) and quark stars. The 
    superdense material of strange quark stars is too expected to be in solid state (Xu 
    2003, 2009).
    \begin{figure}
\centering
 \includegraphics[scale=0.7]{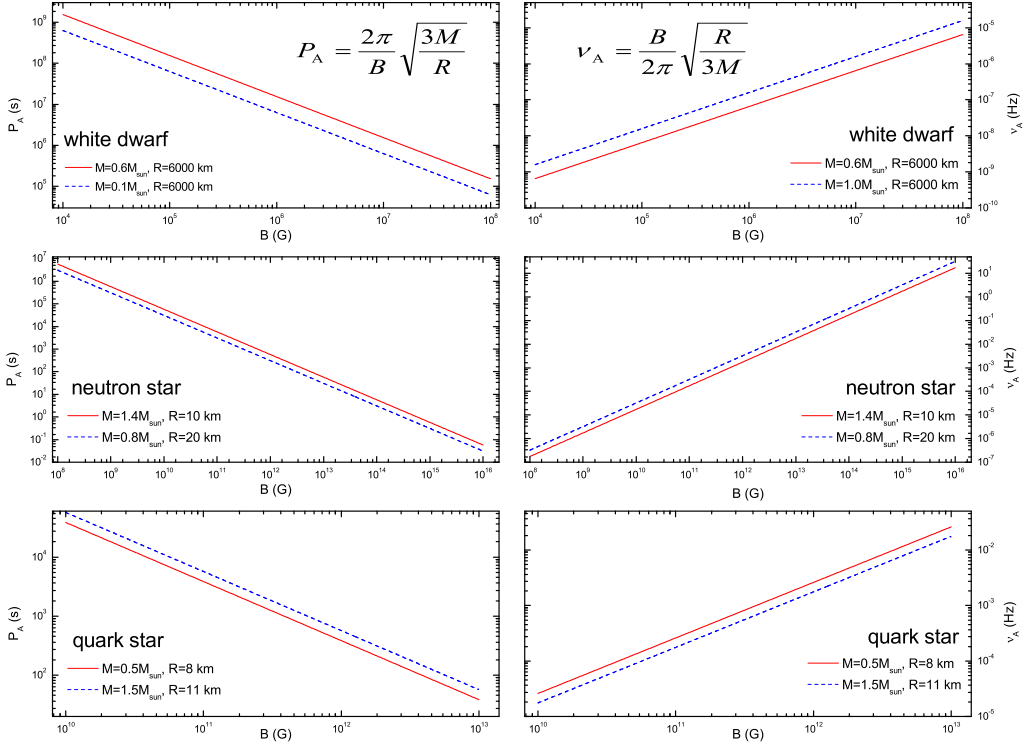}
  \caption{The basic frequencies and periods of global Alfv\'en oscillations as functions of magnetic field for the typical
  mass and radius of white dwarfs, neutron stars and quark stars. The difference in mass and radius can be regarded as reflecting difference in underlying equations of state of matter in these compact objects.}
\end{figure}
 In  Fig.1 we plot 
 the fundamental frequency  $\nu_A=\omega_A/2\pi$
  (where $\omega_A=v_A/R$) and the period $P_A=\nu_A^{-1}$ of global Alfv\'en oscillations
  \begin{eqnarray}
  \label{e2.4}
 \nu_A=\frac{B}{2\pi}\sqrt{\frac{R}{3M}},\quad
 P_A=\frac{2\pi}{B}\sqrt{\frac{3M}{R}}
\end{eqnarray}
as functions of intensity of constant in time magnetic field $B$ for 
solid star models with masses and radii of typical white dwarfs, neutron stars and 
quark stars.

  The total vibration energy and frequency are 
   given by
    \begin{eqnarray}
 \label{e2.5}
  && E_A(t)=\frac{{\cal M}{\dot \alpha}^2(t)}{2}+\frac{{\cal K}(B(t))\alpha^2(t)}
  {2},\quad \omega(t)=\sqrt{\frac{{\cal K}(t)}{\cal M}}=B(t)\kappa,\quad\kappa=\sqrt{\frac{R}{3M}}\,s.
  \end{eqnarray}
  Here $M$ and $R$ are the neutron star mass and radius and $s$ is the parameter
  depending on overtone of Alfven toroidal mode and the depth of seismogenic  
  layer. The energy conversion of above magneto-mechanical vibrations into magneto- 
   dipole radiation which is governed by equation 
   \begin{eqnarray}
 \label{e2.6}
  && \frac{dE_A(t)}{dt}=-{\cal P}(t),\quad {\cal P}(t)=\frac{2}{3c^3}\delta {\ddot {\mbox{\boldmath
  $\mu$}}}^2(t).
  \end{eqnarray}
   The axisymmetric torsional oscillations of matter around magnetic axis of the star 
   are accompanied by fluctuations of total magnetic moment preserving
  its initial (in seismically quiescent state) direction: $\mbox{\boldmath
  $\mu$}=\mu\,{\bf n}={\rm constant}$. The frequency $\omega(t)$ of 
  such oscillations must be the same for both fluctuations of 
  magnetic momenta $\delta {\mbox{\boldmath $\mu$}}(t)$ 
  and  above magneto-mechanical oscillations which are described in terms of 
  $\alpha(t)$, namely
  \begin{eqnarray}
   \label{e2.7}
  && \delta {\ddot {\mbox{\boldmath $\mu$}}}(t)+\omega^2(t)
  \delta {\mbox{\boldmath $\mu$}}(t)=0,\quad {\ddot \alpha}(t)+\omega^2(t){\alpha}(t)=0,\quad \omega^2(t)=B^2(t)
  {\kappa}^2.
  \end{eqnarray}
  This suggests $\delta \mbox{\boldmath $\mu$}(t)=\mbox{\boldmath $\mu$}\,\alpha(t)$. 
  On account of this the equation of energy conversion 
  is reduced to the following law of magnetic field decay  
   \begin{eqnarray}
  \label{e2.8}
  && \frac{dB(t)}{dt}=-\gamma\,B^3(t),\quad
  \gamma=\frac{2\mu^2\kappa^2}{3{\cal M}c^3}={\rm
  constant},\\
   \label{e2.9}
  && B(t)=\frac{B(0)}{\sqrt{1+t/\tau}},\quad \tau^{-1}=2\gamma B^2(0).
  \end{eqnarray}
  Thus, the magnetic field decay resulting in the loss of total energy
   of Alfv\'en vibrations of the star causes its vibration period $P\sim B^{-1}$ 
   to lengthen at a rate proportional to the rate of magnetic field decay. 
   The time evolution of vibration amplitude (the 
  solution of equation for $\alpha(t)$, obeying two 
  conditions $\alpha(t=0)=\alpha_0$ and $\alpha(t=\tau)=0$) reads\footnote{The authors are indebted to Alexey Kudryashov (Saratov, Russia), for the helpful assistance in solving this equation.}(e.g., Polyanin \& Zaitzev 2004) 
   \begin{eqnarray}
    \label{e2.10}
 &&\alpha(t)=C\,s^{1/2}\{J_1(z(t))-\eta\,Y_1
 (z(t))\},\,\,z(t)=2\beta\,s^{1/2}(t),\quad s=1+t/\tau
  \end{eqnarray}
 where $J_1(z)$ and $Y_1(z)$ are Bessel functions (e.g., Abramowitz \& Stegun 1972) 
 and \begin{eqnarray}
\label{e2.11}
 &&\alpha_0^2=\frac{2{\bar E_A(0)}}{M\omega^2(0)}=\frac{2{\bar E_A(0)}}{K(0)},\quad
 \omega^2(0)=\frac{K(0)}{M}.
  \end{eqnarray}
 Here by ${\bar E}_A(0)$ is understood the average energy stored
 in torsional Alfv\'en vibrations of magnetar. If all the detected energy 
 $E_{\rm burst}$ of $X$-ray outburst goes 
  in the quake-induced vibrations, $E_{\rm burst}=E_A$, then the initial 
  amplitude $\alpha_0$ is determined unambiguously.
   \begin{figure}
 \centering
 \includegraphics[scale=0.5]{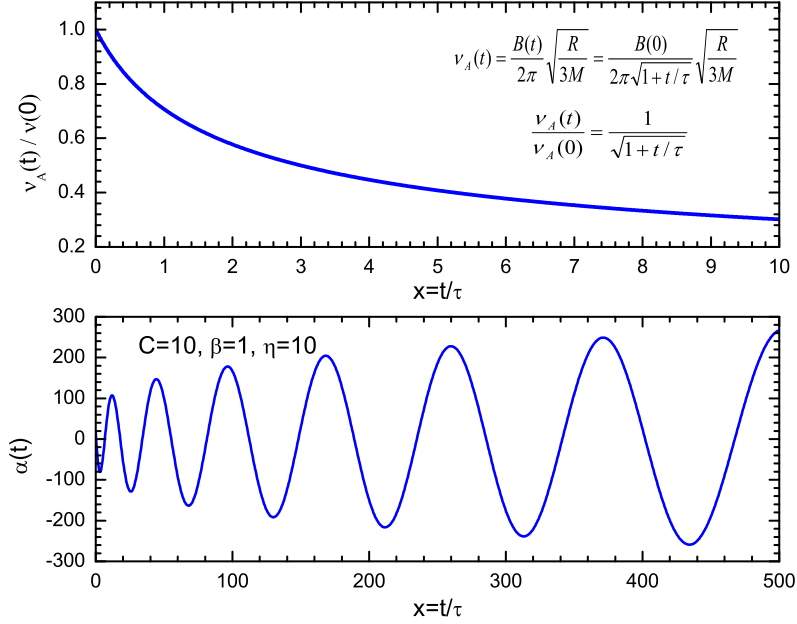}
 \caption{The figure illustrates the effect of magnetic field decay on 
 the vibration frequency and amplitude of quadrupole toroidal $a$-mode presented 
 as functions of $x=t/\tau$.}
 \end{figure}
   Thus, the magnetic field decay is crucial to the energy conversion from Alfv\'en 
   vibrations to electromagnetic radiation whose most striking feature is 
   the  gradual decrease of frequency of pulses (period lengthening). 
    The magnetic-field-decay induced lengthening of period of pulsating radiation  
(equal to period of vibrations) is described by  
   \begin{eqnarray}
 \nonumber
 && P(t)=P(0)\,[1+(t/\tau)]^{1/2},\,\,
 {\dot P}(t)=\frac{1}{2\tau}\frac{P(0)}{[1+(t/\tau)]^{1/2}},\\
 && \tau=\frac{P^2(0)}{2P(t)\dot P(t)},\quad P(0)=\frac{2\pi}{\kappa\,B(0)}
 \label{e2.12}
 \end{eqnarray}
 On comparing $\tau$ given by equations (\ref{e2.9}) and (\ref{e2.12}), one 
 finds that interrelation between equilibrium equilibrium value of the total 
 magnetic moment $\mu$ of a neutron star of mass $M=1.2\,M_\odot$ and radius $R=15$ km vibrating in   
 quadrupole overtone of toroidal $a$-mode, pictured in Fig.3, is given by 
  \begin{eqnarray}
     \label{e2.14}
&& \mu=5.5 \times 10^{37}\,\sqrt{P(t)\,{\dot P}(t)},\,\,{\rm G\,cm^3}.
\end{eqnarray}

\begin{figure}
 \centering
 \includegraphics[scale=0.5]{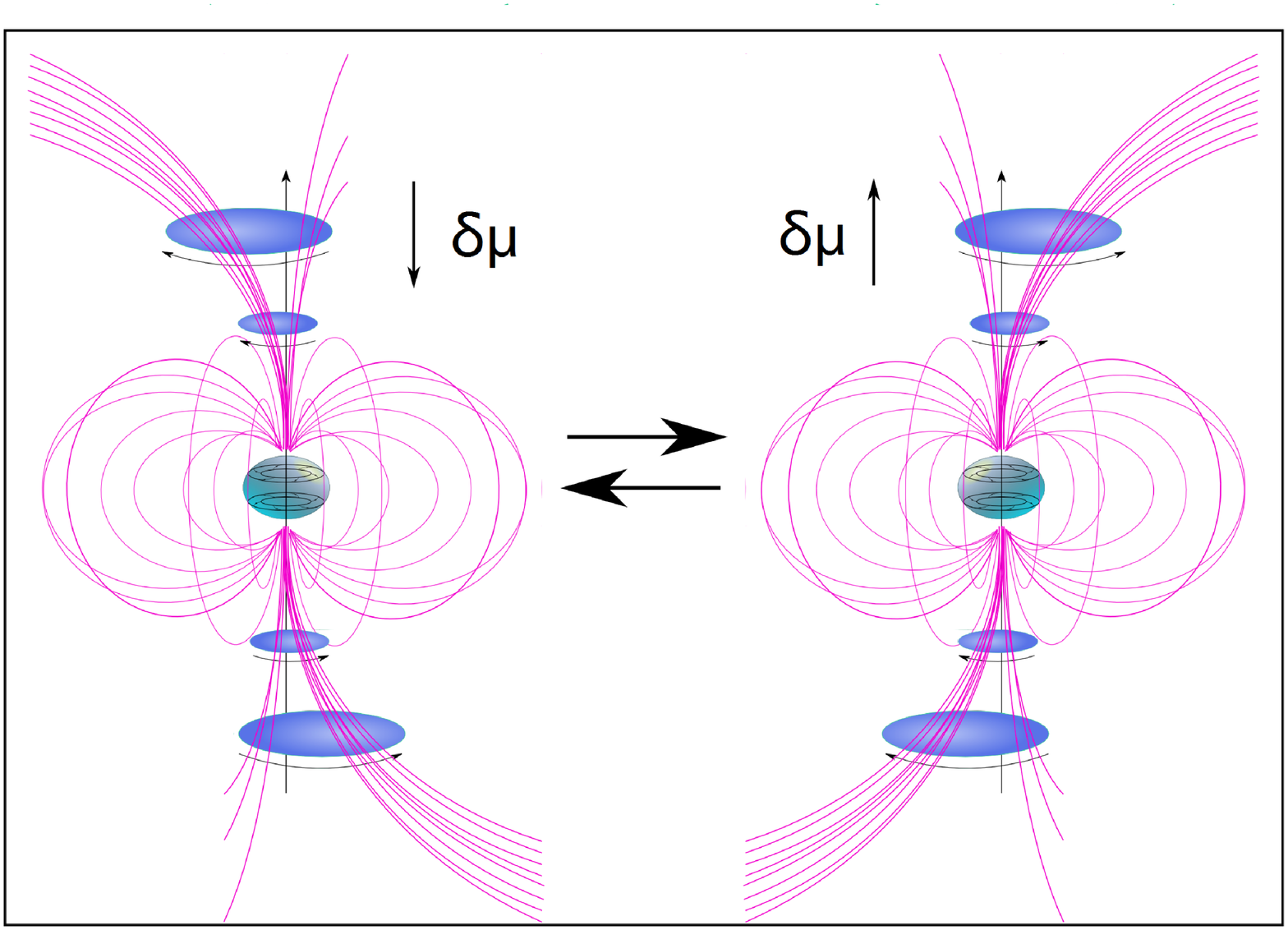}
  \caption{Artist view of quadrupole overtone of torsional Alfv\'en vibrations 
  of magnetar about magnetic axis producing oscillations of lines of dipolar magnetic 
  field defining the beam direction of outburst X-ray emission.}
  \label{fig:fr-period}
\end{figure}
 This illustrative picture shows that 
pulsating radiation in question belongs to the periodic changes of polarization of 
magneto-dipole radiation powered by Alfv\'en seismic vibrations of neutron star. 
In the reminder we present numerical estimates of characteristic parameters of 
of pulsating magneto-dipole radiation produced by quaking neutron star 
at the expense of energy of  Lorentz-force driven differentially rotational vibrations about axis of dipole magnetic moment. 

\section{Numerical analysis}

To get an idea of ​​the magnitude ​​of characteristic parameter of vibrations providing 
energy supply of magneto-dipole radiation of a neutron star, in Table 1 we present results of numerical computations of the fundamental frequency of neutron star oscillations in quadrupole toroidal $a$-mode and time of decay of magnetic field $\tau$ as functions of increasing magnetic field.

 \begin{table}
\begin{center}
\caption{The Alfv\'en frequency of Lorentz-force-driven torsion vibrations, $\nu_A$, and their lifetime equal to decay time of magnetic field, $\tau$, in neutron stars with magnetic fields typical for pulsars and magnetars.}\label{TVT}
  \begin{tabular}{cccccc}
    \hline\hline
    & $M$($M_{\odot}$)  &$ R$(km)  & $B$(G)&  $\nu_{\rm A}$(Hz)  & $\tau$(yr)  \\
    \hline
  Pulsars &0.8 &20 &$10^{12}$ & $3.25\times 10^{-3}$ & $4.53 \times 10^{10}$\\
   & 1.0 & 15 & $10^{13}$ & $2.52 \times 10^{-2}$ &$2.98\times 10^{7}$  \\
  Magnetars &1.1 & 13& $10^{14}$ &0.22 & $7.4 \times 10^{3}$ \\
  &1.2 & 12& $10^{15}$ &2.06  &1.31  \\
  &1.3 & 11& $10^{15}$ &1.89  &2.38  \\
  &1.4 & 10 & $10^{16}$  &17.4  &$4.44 \times 10^{-4}$ \\ 
     \hline
  \end{tabular}\\
\end{center}
\end{table}
As was emphasized, the most striking feature of considered model 
of vibration powered  radiation is the lengthening of periods of pulsating 
emission caused by decay of internal magnetic field. This suggests that this model 
 is relevant to electromagnetic activity of magnetars - neutron stars endowed
 with magnetic field of extremely high intensity the
 radiative activity of which is ultimately related to the magnetic field decay. 
 Such a view is substantiated by estimates of Alfv\'en frequency 
 presented in the table.   For magnetic fields of 
  typical rotation powered radio pulsars, $B\sim 10^{12}$ G, the computed frequency 
  $\nu_A$ is much smaller than the detected frequency of pulses whose origin is 
  attributed to lighthouse effect.  
  In the meantime, for neutron stars with magnetic fields 
  $B\sim10^{14}$ G the estimates of $\nu_A$  are in the realm of 
  observed frequencies of high-energy pulsating emission of soft gamma repeaters 
  (SGRs), anomalous X-ray pulsars (AXPs) and sources exhibiting similar 
  features. According to common belief, these are magnetars - highly magnetized 
  neutron stars whose radiative activity is related with magnetic field decay (e.g., 
  Woods \& Thompson 2006). The amplitude of vibrations is estimated
  as  
\begin{eqnarray}
\label{e3.1}
&& \alpha_{0}= \left[\frac{2\bar E_{\rm A}(0)}{{\cal M}\omega^{2}(0)}\right]^{1/2}=3.423 \times 10^{-3} \bar E_{\rm A, 40}^{1/2} B^{-1}_{14} R_{6}^{-3/2}.
\end{eqnarray}
where $\bar E_{\rm A, 40}= \bar E_{\rm A}/(10^{40}\; {\rm erg})$ is the energy stored in the vibrations. $R_{6} = R/(10^{6} \; {\rm  cm})$ and $B_{14} = B/(10^{14} \; {\rm G})$. 
The presented computations show that the decay time of magnetic 
  field (equal to duration time of vibration powered radiation in question) strongly 
  depends on the intensity of initial magnetic field of the star: the larger 
  magnetic field $B$ the shorter time of radiation $\tau$ at the expense of energy 
  of vibration in decay during this time magnetic field. The effect of equation of state
  of neutron star matter (which is most strongly manifested in different values mass and radius of the star) on frequency $\nu_A$ is demonstrated by numerical vales of this quantity for magnetars with one and the same value of magnetic field $B=10^{15}$ G but different values of mass and radius.

According to above presented analytic results, the neutron star 
luminosity powered by neutron star vibrations in quadrupole toroidal $a$-
is given by
\begin{eqnarray}
\label{e3.1}
&& {\cal P}=\frac{\mu^2}{c^3}\kappa^4 B^4(t) \alpha^2(t),\quad \mu=(1/2)B(0)R^3,\quad B(t)=B(0)[1+t/\tau]^{-1/2},\\
&& \alpha(t)=C\,s^{1/2}\{J_1(z(t))-\eta\,Y_1 (z(t))\},\quad z=2\omega(0)\tau\,(1+t/\tau)
\end{eqnarray} 
The presented in Fig.4 computations of power of magneto-dipole radiation 
of a neutron stars (of one and the same mass and radius but different values of magnetic fields) exhibit oscillating character of luminosity. 
The frequency of these oscillations equal to that of torsional Alfv\'en seismic vibrations of neutron star.  The practical usefulness of presented computations 
 is that they can be used as a guide in observational quest of vibration 
 powered neutron stars among the currently monitoring AXP/SGR-like sources.

\begin{figure}
 \centering
 \includegraphics[scale=0.7]{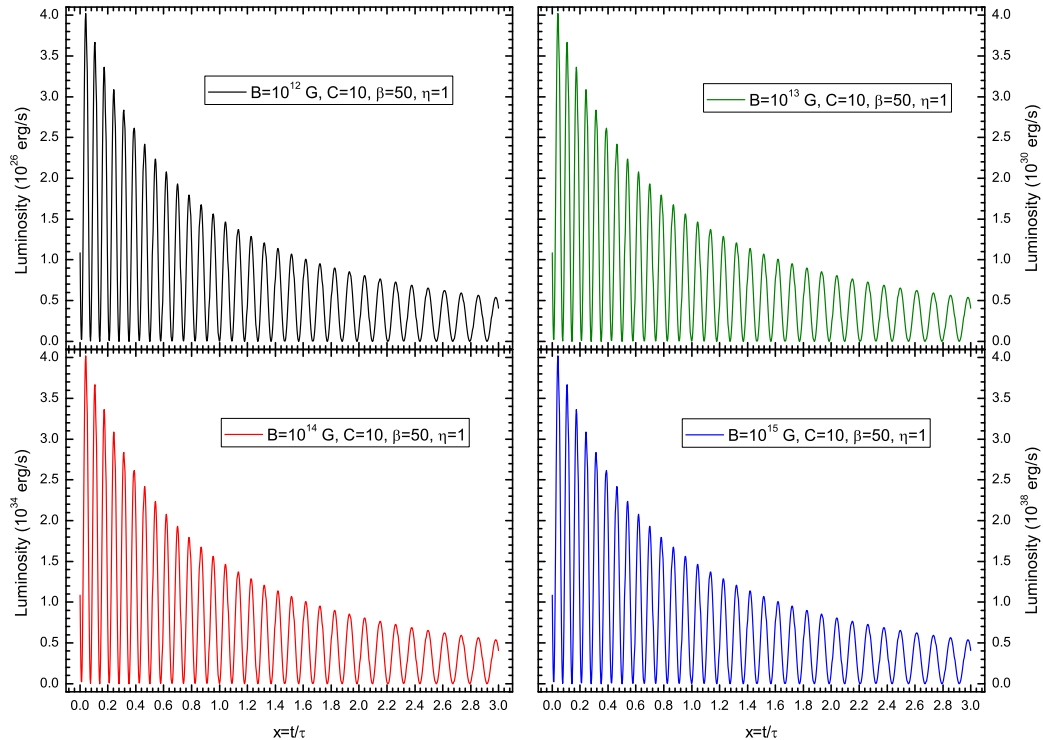}
  \caption{Time evolution of luminosity of magneto-dipole radiation 
  powered by energy of torsional Alfv\'en seismic vibrations 
  of neutron star with mass $M=1.2M_\odot$ and radius $R=15$ km 
  with intensities of magnetic field.}
  \label{fig:fr-period}
\end{figure}

\section{Summary}

It is generally realized today that the standard model of 
 inclined rotator, lying at the base of our understanding of radio pulsars, faces 
 serious difficulties in explaining  the long-periodic ($2<P<12$ s) pulsed radiation of 
 soft gamma repeaters (SGRs) and anamalous $X$-ray pulsars (AXPs). 
 Observations show that persistent $X$-ray luminosity of these sources      
 ($10^{34}<L_X<10^{36}$ erg  s$^{-1}$) is  appreciably (10-100 times) larger 
 than expected from neutron star deriving radiation power from energy of rotation 
 with frequency of detected pulses. It is believed that this discrepancy can be resolved 
 assuming that AXP/SGR-like sources 
 are magnetars -- young, isolated and seismically active neutron stars whose energy supply of pulsating high-energy radiation comes not from 
 rotation (as is the case of radio pulsars) but from different process
 involving decay of ultra strong magnetic field,  $10^{14}<B<10^{16}$ G.
 Adhering to this attitude we have investigated the model of neutron star 
 deriving radiation power from the energy of torsional Lorentz-force-driven oscillations in its own magnetic field experiencing decay. It worth noting that such an idea is not new and first has been discussed by Hoyle et al. (1964), still before the discovery of pulsars (e.g., Pacini 2008). What is newly disclosed 
 here is that such radiation is possible when magnetic field is decayed.
 Since the magnetic field decay is one of the most conspicuous 
 features distinguishing magnetars from rotation powered 
 pulsars,  it seems meaningful to expect that at least some of AXP/SGR - like sources can be vibration powered magnetars.
 Working from this, we have presented extended numerical analysis of the model 
 of vibration powered neutron star whose results can be efficiently, as is hoped, 
 utilized as a guide for discrimination of vibration powered from rotation powered neutron stars. The application of presented here analytic and numerical results to analysis of specific currently monitoring sources will be be presented 
 in forthcoming papers.

\section*{Acknowledgments}
The authors are grateful to anonymous 
 referee for a set of valuable suggestions.  
This work is supported by the National Natural Science Foundation of China
(Grant Nos. 10935001, 10973002), the National Basic Research Program of
China (Grant No. 2009CB824800), and the John Templeton Foundation.

\label{lastpage}

\end{document}